\documentclass[oneside,english,reqno]{amsart}
\usepackage[T1]{fontenc}
\usepackage[latin1]{inputenc}
\usepackage{prettyref}
\usepackage{setspace}
\usepackage{amssymb}

\makeatletter
 \usepackage{verbatim}
 \theoremstyle{plain}
 \theoremstyle{definition}
  \newtheorem{example}{Example}
 \theoremstyle{plain}    
 \newtheorem{prop}{Proposition} 
 \theoremstyle{plain}    
 \newtheorem{thm}{Theorem} 
 \theoremstyle{remark}    
 \newtheorem*{acknowledgement*}{Acknowledgement} 

\usepackage{babel}
\makeatother
\begin{document}

\newcommand{\qgr}{(G,\vartheta)}

\newcommand{\radqgr}{(\sqrt{G},\sqrt{\vartheta})}

\newcommand{\cqgr}{(\mathbb{Z}_{N},\vartheta)}

\newcommand{\sca}{(X,\omega)}

\newcommand{\minsc}[1]{M}

\newcommand{\ca}[2]{\mathcal{W}[#1,#2]}

\newcommand{\aca}[1]{\mathcal{A}[#1]}

\title{Simple Current Actions of Cyclic Groups}

\author{Tamas Varga}

\address{{\small Department of Theoretical Physics,} Roland Eötvös University,
1117 Budapest, Hungary}

\email{vargat@elte.hu}

\begin{abstract}
Permutation actions of simple currents on the primaries of a Rational
Conformal Field Theory are considered in the framework of admissible
weighted permutation actions. The solution of admissibility conditions
is presented for cyclic quadratic groups: an irreducible WPA corresponds
to each subgroup of the quadratic group. As a consequence, the primaries
of a RCFT with an order $n$ integral or half-integral spin simple
current may be arranged into multiplets of length $k^{2}$ (where
$k$ is a divisor of $n$) or $3k^{2}$ if the spin of the simple
current is half-integral and $k$ is odd.
\end{abstract}
\maketitle
\begin{comment}
81T40

conformal field theory, simple currents

Work supported by grant OTKA TS044839
\end{comment}

\section{Introduction}

The construction of correlation functions from conformal blocks is
restricted in a Rational Conformal Field Theory by requiring the invariance
of physical quantities under the action of mapping class groups \cite{ms,frsch1}.
The most important case is that of the torus partition function -
it has to be an invariant of the representation of the modular group
$SL(2,\mathbb{Z})$ on the characters of the theory. The modular representation
is determined by matrices $S$ and $T$ (indexed by the primaries
of the theory) corresponding to modular transformations $\tau\to-1/\tau$
and $\tau\to\tau+1$, respectively. $S$ is symmetric, unitary, its
square is the charge conjugation matrix, and $T$ is a finite order
diagonal matrix. Symmetries of the modular representation turned out
to be the most effective tool in constructing modular invariant partition
functions. One of these symmetries is provided by simple currents
\cite{schy1,schy2,intril,b-sc,halpsc} - the corresponding simple
current modular invariants yield most of the known modular invariant
partition functions \cite{kreuzersch}. 

Simple currents are primary fields whose fusion with any primary field
contains only one term, i.e. whose fusion matrix is a permutation
matrix

\begin{equation}
N_{\alpha p}^{q}=\Pi(\alpha)_{p}^{q}=\delta_{\alpha p}^{q}.\label{eq:scfusmat}\end{equation}
 Other equivalent definitions of simple currents could be used: they
are the primaries whose fusion with their charge conjugate gives the
vacuum alone, or, in unitary theories, the primaries with quantum
dimension $d_{p}=S_{p0}/S_{00}=1$. It follows from the associativity
and commutativity of the fusion ring that they form an abelian group
under fusion, known as the the simple current group, or the center
of the fusion ring. The fusion of simple currents with other primaries
of the theory gives a permutation action of the simple current group
on the set of primaries. 

It is known from Verlinde's theorem \cite{verl} that fusion matrices
are diagonalized by $S$, which applied on \prettyref{eq:scfusmat}
yields

\begin{equation}
S_{\alpha p}^{q}=\theta(q,\alpha)S_{p}^{q},\label{eq:scsymm}\end{equation}
where the complex number $\theta(q,\alpha)$ is the (exponentialized)
monodromy charge of the primary $q$ with respect to the simple current
$\alpha$. The above equation is a symmetry of the $S$ matrix, which
relates its rows in the same orbit of the permutation $\alpha$. In
fact, it can be shown using Verlinde's theorem and unitarity, that
if $\alpha$ is a permutation satisfying \prettyref{eq:scsymm}, then
$\alpha0$ is a simple current, where $0$ denotes the vacuum. Therefore,
these symmetries of the $S$ matrix are in one-to-one correspondence
with simple currents and are called simple current symmetries. 

Simple currents have various other uses besides the already mentioned
simple current modular invariants: they are used in the GSO projection
in string theory \cite{fschw}, through simple current extension they
allow us to construct a new vertex operator algebra and modular representation
\cite{fss,ijmpa} from a known one, and they are also used in the
determination of the projective kernel of modular representation \cite{b-projk}.
In most of these applications the monodromy charges defined in \prettyref{eq:scsymm}
play an important role, but hardly any more properties of the modular
representation are needed. This suggests to consider simple currents
and monodromy charges separately from the modular representation -
that approach lead in \cite{b-sc} to the introduction of weighted
permutation actions. 

The monodromy charges can be expressed using the conformal weights
$\Delta_{p}$ and the central charge of the theory $c$ as follows.
Let $\omega(p)$ denote the diagonal entry of $T$ corresponding to
the primary field $p$\[
\omega(p)=T_{p}^{p}=exp(2\pi i(\Delta_{p}-\frac{c}{24})),\]
 and let $\vartheta(\alpha)=\omega(\alpha)/\omega(0)=exp(2\pi i\Delta_{\alpha}).$
Using \prettyref{eq:scsymm} and the modular relation $STS=T^{-1}ST^{-1}$
one may express $\theta(p,\alpha)$ in terms of $\omega$ as

\begin{equation}
\theta(p,\alpha)=\frac{\omega(p)\omega(\alpha)}{\omega(\alpha p)\omega(0)}=\frac{\omega(p)}{\omega(\alpha p)}\vartheta(\alpha).\label{eq:monodr}\end{equation}
The above equation shows that the monodromies may be replaced by the
weight function $\omega$, which turns out to be more suitable for
the description of simple currents. It follows from \prettyref{eq:scsymm}
that $\theta(p,\alpha\beta)=\theta(p,\alpha)\theta(p,\beta)$, and
substituting the above expression for $\theta(p,\alpha)$ yields\begin{equation}
\frac{\omega(\alpha p)\omega(\beta p)}{\omega(p)\omega(\alpha\beta p)}=\frac{\vartheta(\alpha)\vartheta(\beta)}{\vartheta(\alpha\beta)}.\label{eq:def weight}\end{equation}
 If $p$ is chosen to be a simple current $\gamma$, one obtains the
following equality for $\vartheta$

\begin{equation}
\vartheta(\alpha\beta)\vartheta(\beta\gamma)\vartheta(\gamma\alpha)=\vartheta(\alpha)\vartheta(\beta)\vartheta(\gamma)\vartheta(\alpha\beta\gamma).\label{eq:qform1}\end{equation}
In a unitary theory one may also show that

\begin{equation}
\vartheta(\alpha^{n})=\vartheta(\alpha)^{n^{2}}.\label{eq:qform2}\end{equation}
These two properties define $\vartheta$ to be a \emph{quadratic form}
(written multiplicatively) on the abelian group of simple currents
$G$, and the pair $\qgr$ is called a \emph{quadratic group}. A \emph{weighted
permutation action (WPA)} of a quadratic group $\qgr$ is a pair $\sca$
such that the group $G$ acts by permutations on the finite set $X$,
and the function $\omega:X\mapsto\mathbb{C}^{*}$ satisfies \prettyref{eq:def weight}.
$X$ is called the support of the WPA, $\omega$ its weight function,
and $|X|$ the degree of the WPA. 

These results give a mathematical formalization of simple current
symmetries: the group of simple currents together with $\vartheta=exp(2\pi i\Delta_{\alpha})$
is a quadratic group and its permutation action on the set of primaries
together with $\omega(p)$ is a weighted permutation action of it.
The latter is called the simple current WPA associated to the RCFT.

Weighted permutation actions can be completely classified in terms
of coset WPA-s - the details are given in the next section. Once we
know the classification, it is natural to ask: what characterizes
those WPA-s that are associated to some RCFT. In \cite{b-sc} three
necessary conditions were found, and the WPA-s satisfying these were
called admissible. The solution of the admissibility conditions for
a given quadratic group amounts to finding a finite number of irreducible
WPA-s. This can be done in principle for any quadratic group, but
the complexity of the problem grows rapidly. Therefore, the solution
is known only for the most trivial quadratic groups: e.g. prime order
cyclic groups or some abelian groups of low order. Our aim is to present
the solution for cyclic quadratic groups. Note that not all of these
solutions is realized by some RCFT. Nevertheless, it provides us with
non-trivial information about what the primary field content (including
the differences of conformal weights in the same orbit) and simple
current action \emph{may} be in any RCFT. This would allow us, among
others, to describe the possible modular invariants of cyclic simple
current groups. Moreover, for some WPA-s, the corresponding simple
current symmetry \prettyref{eq:scsymm}, together with non-linear
equations like unitarity or $S^{2}=C$, completely determines the
modular representation. 

In the next section we shall shortly review some results of \cite{b-sc}
in order to formulate the admissibility conditions. In \prettyref{sec:Examples}
we give examples of admissible WPA-s. In particular, Example \prettyref{exa:cosetwpa}
provides a general construction of an irreducible WPA corresponding
to each subgroup of a quadratic group. \prettyref{sec:aca} is devoted
to the proof of \prettyref{thm:cycl irred}, whose immediate consequence
is that in the case of cyclic quadratic groups the irreducibles of
Example \prettyref{exa:cosetwpa} exhaust the set of all irreducible
WPA-s.

\section{Admissible weighted permutation actions}

Let us review some elements of the theory of permutation actions that
may be generalized for WPA-s. A WPA is called transitive if it consist
of a single $G$ orbit. Given two WPA-s $(X_{1},\omega_{1})$ and
$(X_{2},\omega_{2})$, it is straightforward to define their direct
sum, whose support is the set $X_{1}\cup X_{2}$. Two WPA-s $(X,\omega)$
and $(X',\omega')$ are considered equivalent if they are equivalent
as permutation actions, and if their weight functions differ by a
factor which is locally constant on $G$ orbits - note that such rescalings
are allowed by \prettyref{eq:def weight}. 

The radical of $\qgr$ is the quadratic group $\radqgr$ where $\sqrt{G}=\{\alpha\in G|$
$\vartheta(\alpha,\beta)=\vartheta(\alpha)\vartheta(\beta)$ $\forall\beta\in G\}$
and $\sqrt{\vartheta}$ is the restriction of $\vartheta$ to $\sqrt{G}$.
It follows from the fact that $\sqrt{\vartheta}$ is both a character
and a quadratic form on $\sqrt{G}$ that the allowed values of $\sqrt{\vartheta}$
are $\pm1$. $\sqrt{\vartheta}(\alpha)=-1$ is allowed only if $\alpha$
is of even order. 

Transitive WPA-s may be classified up to equivalence as follows. Let
$\xi$ be a character of $\sqrt{G}$ and $H$ a subgroup of $ker(\xi\sqrt{\vartheta})$.
Further, let $X$ be the coset space $X=G/H$ on which $G$ acts by
left translations and $\omega(\alpha H)=\vartheta(\alpha)/\xi^{*}(\alpha)$,
where $\xi^{*}$ is any character of $G$ whose restriction to $\sqrt{G}$
equals $\xi$. Then $(X,\omega)$ is a transitive WPA of $\qgr$ called
a coset WPA, whose equivalence class shall be denoted as $\ca{H}{\xi}$.
The coset WPA is well defined and is determined by $\xi$ - different
choices for $\xi^{*}$ lead to equivalent WPA-s. With this classification
of transitive actions we may always write a WPA $\Phi=(X,\omega)$
as a direct sum of coset WPA-s \begin{equation}
\Phi=\bigoplus_{i\in I}n_{i}\ca{H_{i}}{\xi_{i}}.\label{eq:gen_decomp}\end{equation}
 Therefore, a WPA is determined by the non-negative integer multiplicities
$n_{i}$ up to equivalence. 

The coset WPA $R=\ca{1}{\xi_{0}}$, where 1 is the trivial subgroup
and $\xi_{0}$ is the trivial character of $G$, is called the regular
WPA. The importance of the regular WPA lies in the fact that it represents
the transitive component containing the vacuum in any simple current
WPA.

There is another numerical characterization of WPA-s. To each WPA
we may associate the monomial matrices

\begin{equation}
Y(\alpha,\beta)_{q}^{p}=\vartheta(\beta)\frac{\omega(q)}{\omega(\beta q)}\delta_{p}^{\alpha p},\label{eq:Y repr. def.}\end{equation}
where $p,q\in X$. They satisfy the multiplication rule\begin{equation}
Y(\alpha_{1},\beta_{1})Y(\alpha_{2},\beta_{2})=\frac{\vartheta(\alpha_{2})\vartheta(\beta_{1})}{\vartheta(\alpha_{2}\beta_{1})}Y(\alpha_{1}\alpha_{2},\beta_{1}\beta_{2}),\label{eq:Y mult. rule}\end{equation}
i.e. they form a projective representation of $G\times G$. If the
WPA is associated to some RCFT, then the commutation rule of $Y(\alpha,\beta)$
with $M$ representing the $SL(2,\mathbb{Z})$ element $\left(\begin{array}{cc}
a & b\\
c & d\end{array}\right)$ on the space of genus 1 holomorphic blocks is

\begin{equation}
M^{-1}Y(\alpha,\beta)M=\frac{\vartheta(\alpha)^{b(c-a)}\vartheta(\beta)^{c(b-d)}}{\vartheta(\alpha\beta)^{bc}}Y(\alpha^{a}\beta^{c},\alpha^{b}\beta^{d}).\label{eq:sl(2,z) commut. rule}\end{equation}

The trace of $Y(\alpha,\beta)$ gives a useful numerical description
of the equivalence classes of WPA-s

\begin{equation}
\Upsilon_{\Phi}(\alpha,\beta)=TrY(\alpha,\beta)=\vartheta(\beta)\sum_{p\in Fix_{\Phi}(\alpha)}\frac{\omega(p)}{\omega(\beta p)},\label{eq:chardef}\end{equation}
where $Fix_{\Phi}(\alpha)$ is the subset of the support of $\Phi$
whose elements are fixed by $\alpha$. The value of $\Upsilon_{\Phi}(\alpha,\beta)$
is zero unless $\alpha,\beta\in\sqrt{G}$, so it is in fact a function
on the set $\sqrt{G}\times\sqrt{G}$. For the coset WPA $\Phi=\ca{H}{\xi}$
it is given as\begin{equation}
\Upsilon_{\Phi}(\alpha,\beta)=\left\{ \begin{array}{cc}
\xi(\beta)[G:H] & \textrm{\textrm{if $\alpha\in H$}},\\
0 & \textrm{otherwise.}\end{array}\right.\label{eq:transchar}\end{equation}

For simple current WPA-s this function is related to the so-called
commutator cocycle $\phi_{p}(\alpha,\beta)$ (see \cite{ijmpa}) as
$\Upsilon(\alpha,\beta)=\sum_{p}\phi_{p}(\alpha,\beta)$. This fact,
together with the commutation rule \prettyref{eq:sl(2,z) commut. rule},
implies that simple current WPA-s possess three additional properties,
given in \cite{b-sc}. \emph{Galois invariance} of $\Phi$ means that
in its decomposition \prettyref{eq:gen_decomp} the multiplicities
of transitives $\ca{H}{\xi}$ and $\ca{H}{\xi^{l}}$ are equal for
all $l$ coprime to the exponent of $G$. \emph{Reciprocity} requires
$\Upsilon_{\Phi}(\alpha,\beta)=\Upsilon_{\Phi}(\beta,\alpha)$ for
all $\alpha,\beta\in G$. Finally, \emph{boundedness} is the property
that $|\Upsilon_{\Phi}(\alpha,\beta)|\le|Fix_{\Phi}(\alpha)\cap Fix_{\Phi}(\beta)|$.
WPA-s satisfying Galois invariance, reciprocity and boundedness are
called admissible. 

Let us mention some interesting consequences of the admissibility
conditions. It follows from both Galois invariance or reciprocity
of $\Phi$ that the values of $\Upsilon_{\Phi}$ are integers. Galois
invariance and reciprocity implies that the expression $\vartheta(\alpha)\vartheta(\beta)\Upsilon_{\Phi}(\alpha,\beta)$
depends only on the subgroup $\left\langle \alpha,\beta\right\rangle $
generated by $\alpha$ and $\beta$ (see Proposition \ref{pro:subgroup property}).
Consequently, if $\Phi$ is admissible, then $\Upsilon_{\Phi}$ may
be regarded as an integer valued function on the set of subgroups
of $\sqrt{G}$ generated by at most two elements.

While the admissibility conditions have such non-trivial consequences,
they are also simple enough for practical use - being just linear
equalities and inequalities in terms of the multiplicities $n_{i}$
if $\Phi$ is written as in \prettyref{eq:gen_decomp}. Linearity
implies that the direct sum of admissible WPA-s is again admissible.
Thus, one defines \emph{irreducible} WPA-s as the admissible WPA-s
that cannot be written as a (non-trivial) direct sum of admissible
WPA-s. The problem of finding admissible WPA-s is then reduced to
the problem of finding irreducible WPA-s. Moreover, as another consequence
of the linearity, there is only a finite number of irreducible WPA-s
for any quadratic group $\qgr$. The last important step in the classification
is that the irreducible WPA-s of $\qgr$ are in one-to-one correspondence
with those of $\radqgr$. Therefore, it is enough consider only fully
degenerate quadratic groups, i.e. quadratic groups with $\sqrt{G}=G$.

Although it is principle possible to find irreducible WPA-s of any
quadratic group $\qgr$, in practice the number of irreducible WPA-s
- and thus the length of the computation - grows dramatically with
the number of non-cyclic subgroups of $G$. (A possible explanation
of this is given after Proposition \ref{pro:upsilon cycl =3D fix}.)
Therefore, there are only a few examples where this has been carried
out. In the next section we give a few examples of admissible WPA-s,
which will aid us later in the the solution of admissibility conditions
for cyclic quadratic groups.

\section{Examples\label{sec:Examples}}

In the following $\qgr$ denotes a completely degenerate quadratic
group. Recall that in this case $\vartheta$ is a $\pm1$ valued character
of $G$. Let us also introduce some notation: 1 denotes the trivial
subgroup in any group, $\xi_{0}$ the trivial character, and $R$
the regular WPA $\ca{1}{\xi_{0}}$. 

\begin{example}
Let us first consider admissible WPA-s of lowest possible degree.
The one point transitive $\minsc{\qgr}=\ca{G}{\vartheta}$ has, according
to \prettyref{eq:transchar}, $\Upsilon_{\minsc{\qgr}}(\alpha,\beta)=\vartheta(\beta)$.
The reciprocity condition is satisfied if and only if $\vartheta=\xi_{0}$.
Otherwise, let \begin{equation}
\minsc{\qgr}=\ca{G}{\vartheta}\oplus\ca{ker\vartheta}{\xi_{0}}.\label{eq:min act}\end{equation}
Using \prettyref{eq:transchar} and $[G:ker\vartheta]=2$, one obtains
\[
\Upsilon_{\minsc{\qgr}}(\alpha,\beta)=\left\{ \begin{array}{cc}
3 & \textrm{if }\vartheta(\alpha)=\vartheta(\beta)=1\\
-1 & \textrm{if }\vartheta(\alpha)=\vartheta(\beta)=-1\\
1 & \textrm{if }\vartheta(\alpha)\ne\vartheta(\beta).\end{array}\right.\]

\end{example}
Therefore, $\minsc{\qgr}$ is an admissible WPA of degree 3 (or degree
1 if $\vartheta=\xi_{0}$), and it is called the minimal admissible
WPA of $\qgr$. The admissible WPA associated to the Ising model is
the minimal admissible WPA of $(\mathbb{Z}_{2},\xi_{1})$, where $\xi_{1}$
is the non-trivial character of $\mathbb{Z}_{2}$. However, if $|G|>2$,
the minimal admissible WPA does not contain $R$, and may not correspond
to a simple current WPA. 

\begin{example}
Let us consider the group $G\times\widehat{G}$, where $\widehat{G}$
is the group of homomorphisms from $G$ to $\mathbb{C}^{*}$, and
introduce the natural quadratic form\begin{equation}
\vartheta(\alpha,\phi)=\phi(\alpha).\label{eq:weight of g2 wpa}\end{equation}
 The quadratic group $(G\times\widehat{G},\vartheta)$ is non-degenerate,
$\sqrt{G\times\widehat{G}}$=1, therefore its only transitive WPA
is the regular action $R$ (of degree $|G|^{2}$, with $\omega(\alpha,\phi)=\vartheta(\alpha,\phi)$),
which is consequently admissible. This is, actually, the quadratic
group and the simple current WPA associated to the holomorphic $G$
orbifold model \cite{dvvv}. It is easy to see that $\Upsilon_{R}=0$
except $\Upsilon_{R}((1,\xi_{0}),(1,\xi_{0}))=|G|^{2}$. 

Since we are interested in admissible WPA-s of completely degenerate
quadratic groups, let us restrict the simple current group to the
subgroup $G\simeq G\times\{\xi_{0}\}$. On this subgroup the quadratic
form \prettyref{eq:weight of g2 wpa} becomes trivial, thus we arrive
at the completely degenerate quadratic group $(G,\xi_{0})$. In terms
of the restricted quadratic group, the regular action branches into
orbits of the form $X_{\phi}=G\times\{\phi\}$. It is clear that $X_{\Phi}$
falls in the equivalence class of $\ca{1}{\phi}$, therefore $R$
branches into the following WPA of $(G,\xi_{0})$\begin{equation}
R'=\bigoplus_{\phi\in\widehat{G}}\ca{1}{\phi}.\label{eq:decomp}\end{equation}

\end{example}
Note that $\Upsilon_{R'}(\alpha,\beta)=\Upsilon_{R}((\alpha,\xi_{0}),(\beta,\xi_{0}))=|G|^{2}\delta_{\alpha,1}\delta_{\beta,1}$
since the elements we have removed from the quadratic group had no
fixed points in the regular action, thus did not give a contribution
to $\Upsilon_{R}$ (see \ref{eq:chardef}). Therefore, $R'$ satisfies
the reciprocity and boundedness conditions. Galois invariance is obvious
from \prettyref{eq:decomp}, so we may conclude that $R'$ is an admissible
WPA. Finally, it is easy to see that the direct sum \prettyref{eq:decomp}
defines an admissible WPA even if the quadratic group is $\qgr$,
where $\vartheta$ is any fully degenerate quadratic form.

\begin{example}
\label{exa:cosetwpa}The last example is the combination of the previous
ones into a general form. For a subgroup $H<G$ let $H_{0}$ denote
$H\bigcap ker\vartheta$. As a generalization of \prettyref{eq:min act}
and \prettyref{eq:decomp} let \begin{equation}
\aca{H}=\left(\bigoplus_{\phi\in\widehat{G},H<ker(\vartheta\phi)}\ca{H}{\phi}\right)\oplus\left(\bigoplus_{\phi\in\widehat{G},H_{0}<ker(\phi)}\ca{H_{0}}{\phi}\right),\label{eq:coset wpa}\end{equation}
if $H\neq H_{0}$. If $H=H_{0}$ the two terms in \prettyref{eq:coset wpa}
are equal and only one of them is needed. With this notation Example
1 corresponds to $\aca{G}$ and Example 2 to $\aca{1}$. It can be
shown using \prettyref{eq:coset wpa} and \prettyref{eq:transchar}
that $\Upsilon_{\aca{H}}(\alpha,\beta)=[G:H]^{2}\Upsilon_{\minsc{\qgr}}(\alpha,\beta)$
if $\alpha,\beta\in H$, and is zero otherwise. The boundedness condition
is satisfied by $\aca{H}$ as an equality: $|\Upsilon_{\aca{H}}(\alpha,\beta)|=|Fix(\alpha)\cap Fix(\beta)|$.
Galois invariance can be seen from \prettyref{eq:coset wpa}, therefore
$\aca{H}$ is admissible. It is called the admissible coset WPA corresponding
to the subgroup $H$, since this construction is the simplest way
to extend the coset WPA $\ca{H}{\xi}$ (with arbitrary $\xi$) to
an admissible WPA. The aim of the next section is to prove \prettyref{thm:cycl irred},
which shows the important role played by admissible coset WPA-s.
\end{example}

\section{Admissible coset WPA-s\label{sec:aca}}

Let us first consider properties of the function $\Upsilon_{\Phi}(\alpha,\beta)$
associated to an admissible WPA $\Phi=(X,\omega)$. Admissibility
requires it to be integer valued and symmetric in $\alpha$ and $\beta$
but a stronger requirement can be given. 

\begin{prop}
\label{pro:subgroup property}If $\Phi$ is a WPA of the completely
degenerate quadratic group $\qgr$ satisfying Galois invariance and
reciprocity, then $\vartheta(\alpha)\vartheta(\beta)\Upsilon_{\Phi}(\alpha,\beta)$
depends only on the subgroup $\left\langle \alpha,\beta\right\rangle $
generated by $\alpha$ and $\beta$. 
\end{prop}
\begin{proof}
Reciprocity requires $\vartheta(\alpha)\vartheta(\beta)\Upsilon_{\Phi}(\alpha,\beta)$
to be invariant under $(\alpha,\beta)\to(\beta,\alpha)$. If one takes
into account that $\Upsilon_{\Phi}(\alpha,\beta)$ is non-zero only
for $\alpha,\beta\in\sqrt{G}$, then it is easy to show using \prettyref{eq:chardef}
that $\vartheta(\alpha)\vartheta(\beta)\Upsilon_{\Phi}(\alpha,\beta)$
is also invariant under $(\alpha,\beta)\to(\alpha,\alpha\beta)$.
These transformations together generate any base change in the subgroup
$\left\langle \alpha,\beta\right\rangle $.
\end{proof}
Therefore, to an admissible WPA $\Phi$ we may associate a function
$\Upsilon_{\Phi}(\left\langle \alpha,\beta\right\rangle )=\vartheta(\alpha)\vartheta(\beta)\Upsilon_{\Phi}(\alpha,\beta)$
defined on the subgroups of $\sqrt{G}$ generated by at most two elements.
For the admissible coset action $\aca{H}$ the form of this function
depends on whether $H<ker\vartheta$:\begin{equation}
\Upsilon_{\aca{H}}(K)=\left\{ \begin{array}{ccc}
H<ker\vartheta & H\nless ker\vartheta\\
{}[G:H]^{2} & 3[G:H]^{2} & \textrm{if }K<H_{0}\\
- & [G:H]^{2} & \textrm{if }K\nless H_{0}\textrm{ but }K<H\\
0 & 0 & \textrm{if }K\nless H,\end{array}\right.\label{eq:adm coset char}\end{equation}
where $H_{0}=H\cap ker\vartheta$.

The boundedness condition requires the absolute value of $\Upsilon_{\Phi}$
to be bounded by the fixed point function of the permutation action
corresponding to the WPA. For cyclic subgroups this is a consequence
of reciprocity and Galois invariance, as one may show using Proposition
\ref{pro:subgroup property}.

\begin{prop}
\label{pro:upsilon cycl =3D fix}Let $\Phi$ be a WPA satisfying reciprocity
and Galois invariance. Then $\Upsilon_{\Phi}(\left\langle \alpha\right\rangle )=\vartheta(\alpha)|Fix(\alpha)|$.
\end{prop}
\begin{proof}
It follows from Proposition \ref{pro:subgroup property} that $\Upsilon_{\Phi}(\left\langle \alpha\right\rangle )$
is meaningful. According to \prettyref{eq:chardef}\[
\Upsilon_{\Phi}(\left\langle \alpha\right\rangle )=\vartheta(\alpha)\Upsilon_{\Phi}(\alpha,1)=\vartheta(\alpha)\sum_{p\in Fix(\alpha)}\frac{\omega(p)}{\omega(p)}=\vartheta(\alpha)|Fix(\alpha)|.\]
 
\end{proof}
This Proposition shows that independent inequalities following from
the boundedness condition correspond to non-cyclic subgroups of $\sqrt{G}$
generated by two elements. If we would consider only cyclic quadratic
groups then, according to the previous Proposition, we could omit
the requirement of boundedness in the definition of admissible WPA-s.
This simplifies the problem of finding irreducible admissible WPA-s
to a great extent. However, we shall not restrict ourselves to cyclic
quadratic groups yet, since it is possible to formulate our main result
in a slightly more general setting. We call an admissible WPA $(X,\omega)$
of $\qgr$ cyclic if its stabilizer subgroups $G_{p}<G$ $(p\in X)$
are all cyclic. 

\begin{thm}
\label{thm:cycl irred}Suppose that $\Phi$ is a cyclic admissible
WPA of $\qgr$. Then $\Phi$ is (equivalent to) a direct sum of admissible
coset actions\[
\Phi=\bigoplus_{\alpha\in G}n_{\alpha}\aca{\left\langle \alpha\right\rangle }.\]

\end{thm}
\begin{proof}
Let us consider $\Upsilon_{\Phi}(H)$ for a cyclic admissible $\Phi$.
First, $Fix_{\Phi}(H)=\emptyset$ if $H$ is not cyclic, and boundedness
requires that also $\Upsilon_{\Phi}(H)=0$. Together with Proposition
\ref{pro:upsilon cycl =3D fix} this means that for cyclic admissible
WPA-s the boundedness condition is satisfied as an equality: $|\Upsilon_{\Phi}(H)|=|Fix_{\Phi}(H)|$
$(\forall H<G)$. 

Now, suppose that $K<G$ is a subgroup which is maximal with the property
$\Upsilon_{\Phi}(K)\ne0$, i.e. for any $H>K$ but $H\ne K$ $\Upsilon_{\Phi}(H)=0$.
As we have seen, such a subgroup is necessarily cyclic, let $K=\left\langle \alpha\right\rangle $.
Our aim is to show that then $\Phi=\Phi_{1}\oplus\aca{\left\langle \alpha\right\rangle }$
where $\Phi_{1}$ is admissible (and clearly also cyclic). This would
prove the theorem by induction on the degree of $\Phi$.

Let us consider transitive components of $\Phi$ contributing to $\Upsilon_{\Phi}(\alpha,\cdot)$.
According to \prettyref{eq:transchar} a transitive component $\ca{H_{i}}{\xi_{i}}$
contributes to $\Upsilon_{\Phi}(\alpha,\cdot)$ only if $\alpha\in H_{i}$.
The maximality of $\left\langle \alpha\right\rangle $ and $|\Upsilon_{\Phi}(H)|=|Fix(H)|$
$(\forall H<G)$ implies that if $\alpha\in H_{i}$, then $H_{i}=\left\langle \alpha\right\rangle $.
Thus, we may write $\Phi$ as $\Phi=\Phi_{0}\oplus\Psi$, where $\Psi=\bigoplus_{k\in K}m_{k}\ca{\left\langle \alpha\right\rangle }{\xi_{k}}$
and $\Upsilon_{\Phi_{0}}(\alpha,\beta)=0$ $(\forall\beta\in G)$.
Let us consider $\Psi=\bigoplus_{k\in K}m_{k}\ca{\left\langle \alpha\right\rangle }{\xi_{k}}$.
By the definition of transitive actions $\xi_{k}(\alpha)=\vartheta(\alpha)$
$(\forall k\in K)$. Furthermore, if $\beta\notin\left\langle \alpha\right\rangle $,
then $0=\Upsilon_{\Psi}(\alpha,\beta)=\sum_{k\in K}m_{k}\xi_{k}(\beta)$,
where the first equation follows from the maximality of $\left\langle \alpha\right\rangle $,
and the second from \prettyref{eq:transchar} . This allows us to
express $m_{k}$ using the orthogonality of characters as \[
m_{k}=\sum_{\beta\in G}\Upsilon_{\Psi}(\alpha,\beta)\overline{\xi_{k}(\beta)}=\sum_{\beta\in\left\langle \alpha\right\rangle }\Upsilon_{\Psi}(\alpha,\beta)\overline{\vartheta(\beta)}=:n.\]
Note that $n$ does not depend on $\xi_{k}$, thus \begin{equation}
\Psi=n\bigoplus_{\xi\in\widehat{G},\xi(\alpha)=\vartheta(\alpha)}\ca{\left\langle \alpha\right\rangle }{\xi}.\label{eq:psidecomp}\end{equation}

In the case $\left\langle \alpha\right\rangle \in\ker\vartheta$ the
above is exactly $\Psi=n\aca{\left\langle \alpha\right\rangle }$,
and it only remains to show that $\Phi_{0}$ is admissible. Since
$\Phi$ and $\Psi$ satisfy reciprocity and Galois invariance, also
does $\Phi_{0}$. It follows from the inequality $|\Upsilon_{\Phi_{0}}(K)|\le|\Upsilon_{\Phi}(K)|+|\Upsilon_{\Psi}(K)|$
that $\Upsilon_{\Phi_{0}}(K)=0$ if $K$ is not cyclic. For cyclic
subgroups the boundedness condition is satisfied by Proposition \ref{pro:upsilon cycl =3D fix},
therefore $\Phi_{0}$ is admissible. 

It remains to finish the proof for $\left\langle \alpha\right\rangle \notin\ker\vartheta$.
In that case \begin{equation}
n\aca{\left\langle \alpha\right\rangle }=\Psi\oplus\left(\bigoplus_{\xi\in\widehat{G},\xi(\alpha)=1}n\ca{ker\vartheta}{\xi}\right)=\Psi\oplus\Psi_{1}.\label{eq: decomp1}\end{equation}
 Therefore, we have to show that $\Phi_{0}=\Phi_{1}\oplus\Psi_{1}$,
with $\Psi_{1}$ as above and $\Phi_{1}$ admissible. The argument
is similar as before. Note that we may choose $\alpha$ such that
$\ker\vartheta\cap\left\langle \alpha\right\rangle =\left\langle \alpha^{2}\right\rangle $
(recall that $\vartheta$ is a $\pm1$ valued character). We may repeat
the argument given at the beginning of the proof to show that the
only transitive components of $\Phi_{0}$ contributing to $\Upsilon_{\Phi_{0}}(\alpha,\alpha^{2})$
are of type $\ca{\left\langle \alpha^{2}\right\rangle }{\xi_{k}}$
and $\xi_{k}(\alpha^{2})=\vartheta(\alpha^{2})=1$, so either $\xi_{k}(\alpha)=1$
or $\xi_{k}(\alpha)=-1$. Similarly as before, it follows from $\Upsilon_{\Phi_{0}}(\alpha^{2},\beta)=0$
$(\forall\beta\notin\left\langle \alpha\right\rangle )$ that \begin{equation}
\Phi_{0}=\Phi_{1}\oplus\left(\bigoplus_{\xi\in\widehat{G},\xi(\alpha)=-1}k\cdot\ca{\left\langle \alpha^{2}\right\rangle }{\xi}\right)\oplus\left(\bigoplus_{\psi\in\widehat{G},\psi(\alpha)=1}l\cdot\ca{\left\langle \alpha^{2}\right\rangle }{\psi}\right).\label{eq:decomp2}\end{equation}

Note, by comparing \prettyref{eq: decomp1} and \prettyref{eq:decomp2}
that what we want to show is $l\ge n$. In order to prove this we
have to consider the reciprocity condition for $\Upsilon_{\Phi}(\alpha^{2},\alpha)$.
It follows from \prettyref{eq:psidecomp} that $\Upsilon_{\Psi}(\alpha^{2},\alpha)-\Upsilon_{\Psi}(\alpha,\alpha^{2})=-2n[G:\left\langle \alpha\right\rangle ]^{2}$.
Then, the reciprocity of $\Phi$ implies that $\Upsilon_{\Phi_{0}}(\alpha^{2},\alpha)-\Upsilon_{\Phi_{0}}(\alpha,\alpha^{2})=\Upsilon_{\Phi_{0}}(\alpha^{2},\alpha)=2n[G:\left\langle \alpha\right\rangle ]^{2}$.
With the use of \prettyref{eq:decomp2} and $\Upsilon_{\Phi_{1}}(\alpha^{2},\alpha)=0$
we get $\Upsilon_{\Phi_{0}}(\alpha^{2},\alpha)=2(l-k)[G:\left\langle \alpha\right\rangle ]^{2}$.
Since $l$ and $k$ are positive this implies $l\geq n$, which means
that \[
\Phi=\Phi_{2}\oplus n\cdot\aca{\left\langle \alpha\right\rangle }.\]
We may finish the proof , by showing (similarly as before) that $\Phi_{2}$
is admissible.
\end{proof}
The above theorem proves the irreducibility of cyclic admissible coset
actions as follows. Suppose, that in contrary $\aca{\left\langle \alpha\right\rangle }=\Phi_{1}\oplus\Phi_{2}$,
where $\Phi_{1}$ and $\Phi_{2}$ are admissible. Then either $\Phi_{1}$
or $\Phi_{2}$ satisfies the conditions of the theorem, with $\left\langle \alpha\right\rangle $
as a maximal subgroup, therefore equals $\aca{\left\langle \alpha\right\rangle }$.
It also proves that the only irreducible admissible WPA-s among the
cyclic WPA-s are the cyclic admissible coset WPA-s. The irreducibility
of non-cyclic admissible coset WPA-s may be proven with a slight modification
of the argument used in the proof. 

As the most important application of \prettyref{thm:cycl irred},
let us consider a completely degenerate cyclic quadratic group $(\mathbb{Z}_{n},\vartheta)$.
According to \prettyref{thm:cycl irred} its admissible WPA-s are
of the form:

\begin{equation}
\Phi=\bigoplus_{l|n}m_{l}\aca{H_{l}},\label{eq:cycl quad irr}\end{equation}
where we denoted by $H_{l}$ the order $l$ subgroup of $\mathbb{Z}_{n}$
( where $l$ divides $n$). 

The admissible coset actions in \prettyref{eq:coset wpa} depend on
the completely degenerate quadratic form $\vartheta$. This is determined
by its value on any generator of $\mathbb{Z}_{n}$, let $\alpha$
denote any of them. If $\vartheta(\alpha)=1$, i.e. $\alpha$ is an
integer spin simple current, then $\vartheta$ is the trivial character.
In this case the admissible coset WPA $\aca{H_{l}}$ is of degree
$(n/l)^{2}$. Note that the only irreducible containing the regular
WPA $R$ is $\aca{1}$ (of degree $n^{2}$), which is Example 2 of
\prettyref{sec:Examples}. 

The other possibility is $\vartheta(\alpha)=-1$, i.e. $\alpha$ is
a half-integer spin simple current. This implies that $\alpha$ is
of even order, $n=2k$, and $ker\vartheta=H_{k}$. In that case the
admissible coset action $\aca{H_{l}}$ is of degree $(n/l)^{2}$ if
$H_{l}<H_{k}=ker\vartheta$, i.e. if $l$ divides $k$. However, if
$l$ does not divide $k$ (which means that $n/l$ is odd) the admissible
coset WPA $\aca{H_{l}}$ is of degree $3(n/l)^{2}$ (see \prettyref{eq:adm coset char}).
The regular WPA still appears only in $\aca{1}$, except if the quadratic
group is $(\mathbb{Z}_{2},\xi_{1})$ when the order 3 admissible coset
WPA $\aca{\mathbb{Z}_{2}}$ contains it as well.

In the general case, when $G$ is not a cyclic group, the admissible
coset actions corresponding to its subgroups give only a part of its
irreducible WPA-s. In that case we may apply the result obtained for
cyclic groups if we restrict the quadratic group to any of its cyclic
subgroups. Let us illustrate this on an example. Let $\qgr$ be a
(non-cyclic) fully degenerate quadratic group of order $N$ and let
the exponent of $G$ be $k$. Then there exists an order $k$ cyclic
subgroup of $G$ - let us denote any of these by $H$. The primaries
of any RCFT whose simple current group is $G$ have to contain a $G$
orbit equivalent to the regular WPA $R=\qgr$. What can we say about
admissible WPA-s containing $R$ in this general case? If we restrict
the simple current group to $(H,\vartheta|_{H}),$ then $R$ branches
into $[G:H]=N/k$ copies of the regular WPA $R'=(H,\vartheta|_{H})$.
The only irreducible of $H$ containing $R'$ is the coset WPA $\aca{1}$
(except if $(H,\vartheta|_{H})=(\mathbb{Z}_{2},\xi_{1})$ - we shall
deal with that case later), which is of order $k^{2}$. Therefore,
any admissible WPA of $\qgr$ which contains the regular action should
contain, if restricted to $(H,\vartheta|_{H})$, $N/k$ copies of
$\aca{1}$, thus it has to be at least of degree $Nk$. Finally, let
us consider the case $(H,\vartheta|_{H})=(\mathbb{Z}_{2},\xi_{1})$
- then the exponent of $G$ is 2, so $G=\mathbb{Z}_{2}^{n}$, but
if $n\ge2$ it is possible to choose $H$ to be a subgroup on which
$\vartheta$ is the trivial character. Therefore, the only exception
is $G=(\mathbb{Z}_{2},\xi_{1})$.

\section{Summary}

Quadratic groups and their weighted permutation actions are the mathematical
objects corresponding to simple current symmetries in RCFT. WPA-s
of an arbitrary quadratic group can be classified as a direct sum
of transitive coset actions. Those WPA-s that are associated to simple
currents have to satisfy additional constraints: Galois- invariance,
reciprocity and boundedness. To solutions of these conditions, i.e.
admissible WPA-s of a given (completely degenerate) quadratic group,
are generated by a finite number of irreducible WPA-s. In \prettyref{sec:Examples}
we have considered some examples of irreducible WPA-s - the first
was motivated by the Ising model, the second by holomorphic orbifold
models of abelian groups. The generalization of these, given in Example
\ref{exa:cosetwpa}, leads to a construction that extends a transitive
coset WPA to an irreducible WPA - this way one obtains an irreducible
WPA corresponding to each subgroup of the quadratic group (see \prettyref{eq:coset wpa}).
Propositions \ref{pro:subgroup property} and \ref{pro:upsilon cycl =3D fix}
implied that the number of independent inequalities corresponding
to the boundedness condition equals the number of (non-cyclic) subgroups
of the quadratic group generated by two elements. This suggests that
the simplest case is that of cyclic quadratic groups, or more generally,
cyclic admissible WPA-s. \prettyref{thm:cycl irred} states that these
are in fact generated by admissible coset WPA-s constructed in Example
\ref{exa:cosetwpa}. As a special case, the admissible WPA-s of completely
degenerate cyclic quadratic groups can be given in the form \prettyref{eq:cycl quad irr}.
This result can be applied whenever the simple current group of the
RCFT contains an order $n$ simple current of integral or half-integral
spin to arrange the primaries of the theory in multiplets that in
general consist of several simple current orbits. As we have discussed
after \prettyref{eq:cycl quad irr} the length of these multiplets
is $k^{2}$ (where $k$ is any divisor of $n$) if the simple current
is of integer spin or $k$ is even, and $3k^{2}$otherwise. Our results
can be applied even if the quadratic group is not cyclic, through
considering branching rules to some cyclic subgroup. We illustrated
this on the orbit corresponding to the vacuum, which led to a lower
bound of the number of primaries $Nk$, where $N$ is the order and
$k$ the exponent of the radical of the simple current group. Note
that a degree $N^{2}$ admissible WPA always exists - it is the admissible
coset WPA $\aca{1}$.

\begin{acknowledgement*}
I would like to thank P. Bantay for discussions.
\end{acknowledgement*}

\end{document}